\documentclass[letterpaper,twocolumn, prb, showkeys, aps]{revtex4-1}
\usepackage{graphicx}
\usepackage{amsmath}
\usepackage{bm}

\usepackage[x11names,svgnames]{xcolor}

\DeclareGraphicsExtensions{.eps,.ps,.pdf,.png,.jpg}
\usepackage[colorlinks]{hyperref}
\hypersetup{
    colorlinks=true,
    linkcolor=magenta,
    citecolor=blue,
    filecolor=black,
    urlcolor=blue,
}

\newcommand\er{\mathrm{e}}

\newcommand\kv{\mathbf{k}}
\newcommand\Kv{\mathbf{K}}
\newcommand\Mv{\mathbf{M}}
\newcommand\Qv{\mathbf{Q}}
\newcommand\kvt{\mathbf{\tilde k}}
\newcommand\Gv{\mathbf{G}}
\newcommand\Sv{\mathbf{S}}
\newcommand\Vv{\mathbf{V}}
\newcommand\Tr{\,\mathrm{Tr}\,}

\newcommand\Sigmav{\bm{\Sigma}}
\newcommand\Xiv{\bm{\Xi}}
\newcommand\Bh{\mathbf{\hat B}}

\begin{document}
\title{Topological phases of the Kitaev-Hubbard Model at half-filling}
\author{J. P. L. Faye}
\author{D. S\'en\'echal}
\affiliation{D\'epartment de physique and RQMP, Universit\'e 
 de Sherbrooke, Sherbrooke,  Qu\'ebec, Canada J1K2R1}

\author{S. R. Hassan}
\affiliation{The Institute of Mathematical Sciences, C.I.T. Campus, Chennai 600 113, India}

\date{\today}

\begin{abstract}
The Kitaev-Hubbard model of interacting fermions is defined on the honeycomb lattice and, at strong coupling, interpolates between the Heisenberg model and the Kitaev model.
It is basically a Hubbard model with ordinary hopping $t$ and spin-dependent hopping $t'$.
We study this model in the weak to intermediate coupling regime, at half-filling, using the Cellular Dynamical Impurity Approximation (CDIA), an approach related to Dynamical Mean Field Theory but based on Potthoff's variational principle.
We identify four phases in the $(U,t')$ plane: two semi-metallic phases with different numbers of Dirac points, an antiferromagnetic insulator, and an algebraic spin liquid.
The last two are separated by a first-order transition.
These four phases all meet at a single point and could be realized in cold atom systems.
\end{abstract}
\maketitle

\section{Introduction}
Mott insulators are systems that should be metals within band theory, but are in fact insulators because of  electron-electron interactions.\cite{Mott:1968fk, Imada:1998rp} 
However, the Mott phase is often hidden behind a magnetically ordered phase at low temperature.\cite{Hassan:2013uq}
Spin liquids are non-magnetic Mott-insulators, without broken lattice symmetry, stabilized purely by quantum effects.\cite{fazekas1974} 
In addition to a spectral gap, they are characterized by spin correlations that decay either exponentially, or as a power law, in the case of algebraic spin liquids.\cite{Rantner2001,Affleck1988} 
Experimentally, a spin liquid ground state has been suggested in the organic material $\kappa$-(BEDT-TTF)$_2$Cu$_2$(CN)$_3$,\cite{Shimizu:2003uq} in other systems like $\rm YMnO_3$\cite{Park:2003uq} and, more recently, in materials with a kagome lattice structure.\cite{Helton:2007ys,Lee:2007kx}
Theoretically, spin liquid phases were found, for instance, in the spin-$\frac12$ Heisenberg model on the kagome lattice,\cite{Hermele:2008vn,Yan:2011fk,Iqbal:2013zr} and in the intermediate-coupling Hubbard model on a triangular lattice.\cite{Sahebsara:2008fv}

Tikhonov et $al$.\cite{Tikhonov} have shown that an algebraic spin liquid is realized when a special type of perturbation is added to the Kitaev spin model.\cite{Kitaev:2006fk}
It can be shown that the stability of the spin liquid phase, in that system, is due to time-reversal symmetry.
The existence of an algebraic spin liquid in a model of interacting fermions, the Kitaev-Hubbard model, was shown by Hassan et $al$.\cite{Hassan:2013fk} 
The phase diagram of this model was investigated using the variational cluster approximation (VCA) which allowed the authors to identify a semi-metallic phase, a N\'eel phase and an algebraic spin liquid phase.
 
In this work, we refine the analysis of Ref.~\onlinecite{Hassan:2013fk} by using the cluster dynamical impurity approximation (CDIA). 
This method is more accurate in its treatment of the Mott transition, which appears clearly as a discontinuous transition with hysteresis.
We also reveal a topological transition within the semi-metallic region, between a phase with eight distinct Dirac points, and another phase with only two Dirac points; this is a Lifshitz transition, that carries into the interacting region.
These four phases (the algebraic spin liquid, the antiferromagnet, and the two semi-metallic phases) meet at a single point in the phase diagram.

The paper is organized as follows. We review the model and describe its non-interacting solution in Section~\ref{sec:model}. In Section~\ref{sec:interactions}, we review the methods used in the interacting case (the CDIA), before presenting and discussing our results in Section~\ref{sec:results}.

\section{The non-interacting limit}\label{sec:model}

We will mostly follow the notation of Ref.~\onlinecite{Hassan:2013kx}.
The Kitaev-Hubbard model, defined on the honeycomb lattice, has the following Hamiltonian:
\begin{equation}\label{eq:H}
 H =\sum_{\langle i,j\rangle_\alpha} \left\{c^{\dag}_{i}\left(\frac{t+t'\sigma^{\alpha}}{2}\right) c_{j}+{\rm H.c.}\right \} 
+U \sum_in_{i\uparrow}n_{i\downarrow}
\end{equation}
where $c_{i\sigma}$ annihilates a fermion of spin $\sigma$ at site $i$ (the spin index 
is implicit in the above), $\sigma^\alpha$ are the Pauli matrices
$(\alpha = x, y,z)$, $U$ the Coulomb repulsion for two electrons of opposite spin on 
the same site, $n_{i\sigma} = c^{\dag}_{i\sigma}c_{i\sigma}$ is the number of electrons of spin $\sigma$ at site $i$, and $\langle i,j\rangle_\alpha$ denotes
the nearest-neighbor pairs in the three hopping directions
of the cluster system (see Fig.~\ref{fig:h4-6b} below).
Throughout this work we will express energies relative to $t$, i.e., we will set $t=1$.

In the limit where $U\gg t,t'$, and at half-filling, the Hamiltonian (\ref{eq:H}), 
becomes equivalent to a combination of the Heisenberg and Kitaev\cite{Kitaev:2006fk} Hamiltonians:
\begin{equation}\label{heff2} 
H^{(2)} = \sum_{\langle ij\rangle_\alpha} \Big[
\frac{(1-t'^2)}{U}\Sv_{i}\cdot\Sv_{j}+\frac{2t'^2}{U} S_{i}^\alpha
S_{j}^\alpha\Big]
\end{equation}
Time reversal is applied by changing the sign of the Pauli matrices ($\sigma^\alpha\to-\sigma^\alpha$).
The Hamiltonian (\ref{eq:H}) breaks time-reversal symmetry explicitly when $t'\ne0$, but is invariant under parity (as defined by the exchange of sublattices A and B on the honeycomb lattice).

In the non-interacting limit ($U=0$), the Hamiltonian~\eqref{eq:H}, after Fourier transform, can be expressed in terms of the destruction operators $c_{\kv,A}$ and $c_{\kv,B}$ on the $A$ and $B$ sublattices (the spin index is, again, implicit):
\begin{equation} \label{Eq:H0}
H_0=\sum_{\kv}
  \begin{pmatrix}c^{\dagger}_{\kv,A} & c^{\dagger}_{\kv,B}\end{pmatrix}  \begin{pmatrix} 0 & \Xiv(\kv) \\ \Xiv^{\dagger}(\kv) & 0 \end{pmatrix} 
  \begin{pmatrix}  c_{\kv,A}\\ c_{\kv,B} \end{pmatrix}
\end{equation}
where  $\Xiv(k)=P_3 + P_1e^{ik_2}+P_2e^{-ik_1}$ is a $2\times2$ matrix acting in spin space, with the projectors
$P_{\alpha}=\frac{1}{2}(1+t'\sigma^{\alpha})$ and $k_{1(2)}=\kv\cdot\mathbf{e}_{1(2)}$. 
The vectors $\mathbf{e}_{1(2)}$ are a Bravais basis of the honeycomb lattice: $\mathbf{e}_{1(2)}=(\pm\frac{3}{2}, \frac{\sqrt{3}}{2})$.

For a given $\kv$, the four eigenvalues of $H_0$ are $\pm \varepsilon_{\pm}(\kv)$ with
$\varepsilon^2_{\pm}(\kv)=\frac12\left[\xi(\kv)\pm t'|\textbf{B}(\kv)|\right]$.
We have defined
\begin{equation}
	\xi(\kv) = \frac32(1+t^{\prime2}) + \cos k_1 + \cos k_2 +\cos k_3
\end{equation}
and the vector $\mathbf{B}$ with components
\begin{align}
B_1(\kv) &= 1-t'\sin k_1+\cos k_2+\cos k_3 \nonumber\\
B_2(\kv) &= 1+\cos k_1-t'\sin k_2+\cos k_3 \nonumber\\
B_3(\kv) &= 1+\cos k_1+\cos k_2-t'\sin k_3 
\end{align}
with $k_3 = -k_1-k_2$.
The four-component eigenvector associated with eigenvalue $p\varepsilon_{p'}(\kv)$ ($p'=\pm$ and $p=\pm$) will be denoted $\Phi^{pp'}$.

\subsection{Dirac points}

\begin{figure}[ht]
\begin{center} 
\includegraphics[scale=1]{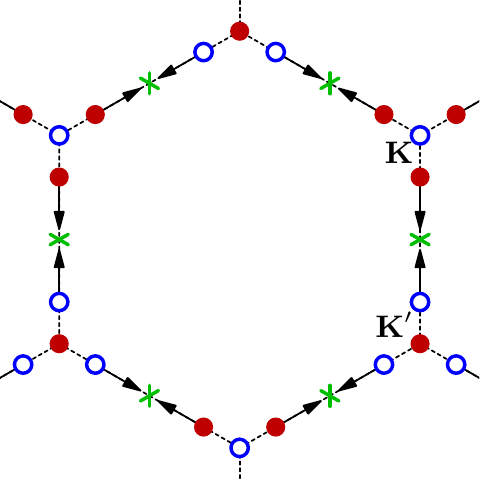}
\caption{Position and chirality of the Dirac points at $U=0$: open (blue) and filled (red) circles are the center of positive and negative circulations of $\nabla\chi$, respectively. As $t'$ increases from zero, the new Dirac points drift in the direction indicated by the arrows, until they annihilate at the points marked by crosses, at the critical value $t'_c$, while the graphene Dirac points stay fixed.
The hexagon is the Brillouin zone.}      
\label{fig:dirac}
\end{center}
\end{figure}

At $t'=0$ (the graphene limit), the Fermi surface consists of two distinct Dirac points $\Kv=(2\pi/3,~2\pi/3\sqrt{3})$ and $\Kv^{\prime}=(2\pi/3,~-2\pi/3\sqrt{3})$ that are the focal points of Dirac cones at positive and negative energies.
As soon as $t'>0$, a total of six new distinct Dirac points appear, along the lines that join $\Kv$ and $\Kv'$, i.e., on the Brillouin zone boundary (Fig.~\ref{fig:dirac}).
These Dirac points form the Fermi surface at half-filling.

The positions of the Dirac points can be found by solving  the equation $\varepsilon_{-}(k_x=\frac{2\pi}3,k_y)=0$ for $k_y$.
In terms of $\Lambda=\cos(\frac{\sqrt{3}}{2}k_y)$, this is a quadratic equation:
\begin{equation} \label{eq:lambda}
4[\Lambda^2(1-t'^2)+\Lambda] +1+t'^2=0
\end{equation} 
whose solutions are
\begin{equation} \label{Eq:sol-lambda}
\Lambda_1=-\frac{1}{2} ~~ \text{and}~~ \Lambda_2=-\frac{1+t'^2}{2(1-t'^2)}
\end{equation}
$\Lambda_1$ corresponds to the graphene Dirac point at the zone corner, whereas $\Lambda_2$ is an additional Dirac point for a given $t'$; the other six Dirac points can be deduced from Eq.~(\ref{Eq:sol-lambda}) by lattice symmetries. The limiting case $\Lambda_2=-1$, for $t'< 1$, corresponds to a critical spin-dependent hopping $t'_c=1/\sqrt{3}$. At that value of $t'$ the six additional Dirac points merge pairwise and disappear at the midpoints between zone corners, as illustrated on Fig.~\ref{fig:dirac}.
This merging is discussed in more detail below.

Note that the Dirac points are protected by parity.
Adding a parity-non-conserving term, such as a staggered magnetization $M$, would change the Hamiltonian \eqref{Eq:H0} into
\begin{equation} \label{eq:M0}
\sum_{\kv}
  \begin{pmatrix}c^{\dagger}_{\kv,A} & c^{\dagger}_{\kv,B}\end{pmatrix}  \begin{pmatrix} M & \Xiv(\kv) \\ \Xiv^{\dagger}(\kv) & -M \end{pmatrix} 
  \begin{pmatrix}  c_{\kv,A}\\ c_{\kv,B} \end{pmatrix}
\end{equation}
and the corresponding energies would become
$\varepsilon^2_{\pm}(\kv)=\frac12\left[\xi(\kv) + M^2 \pm t'|\textbf{B}(\kv)|\right]$, which never vanishes.
Thus all Dirac points disappear if $M\ne0$.

\subsection{The Pancharatnam-Berry curvature}

The Dirac points are singular points of the Pancharatnam-Berry (PB) curvature.
The latter is given by
\begin{equation}
R^{pp'}(\kv)=\frac{\epsilon^{\mu\nu}}{4\pi i}
\partial_\mu\Phi^{pp'}(\kv)^\dagger\partial_\nu\Phi^{pp'}(\kv)
\end{equation}
where $\mu,\nu=1,2$ are two orthogonal directions in the Brillouin zone,
$\epsilon^{\mu\nu}$ is the two-dimensional Levi-Civita tensor, and the eigenstates of the Hamiltonian (\ref{Eq:H0}) are
\begin{equation}\label{eq:eigenvectors}
\Phi^{pp'}(\kv)=\frac{1}{\sqrt{2}}
\begin{pmatrix} \psi^{p'}(\kv) \\ p \er^{i\chi(\kv)}\psi^{p'}(-\kv) \end{pmatrix}
\end{equation}
with 
\begin{equation}
\psi^\pm(\kv)= C(\kv)
\begin{pmatrix} 
B_3(\kv)\mp|\mathbf{B}(\kv)| \\ 
-B_1(\kv) + i B_2(\kv)
\end{pmatrix}
\end{equation}
and $C(\kv)$ is a normalization factor.
The expression for the phase $\chi(\kv)$ is known analytically for all values of $t'$ but is to complex to reproduce here.

It can be shown that
\begin{equation}\label{eq:PB}
R^{pp'}(\kv) = p'\frac{1}{2}\left[b(\kv)+b(-\kv)\right]
+\frac{1}{2}\epsilon^{\mu\nu}\partial_\mu\partial_\nu\chi(\kv)
\end{equation}
where
\begin{equation}
b(\kv) = \frac{\epsilon^{\mu\nu}}{8\pi}
\Bh(\kv)\cdot\partial_\mu\Bh(\kv)\times\partial_\nu\Bh(\kv)
\end{equation}
The first term of \eqref{eq:PB} is everywhere regular, but the last term is singular where the phase $\chi$ is ill-defined, which occurs when two bands cross at a given wave vector, i.e., at Dirac points.

\begin{figure}[ht]
\begin{center} 
\includegraphics[width=0.4\textwidth]{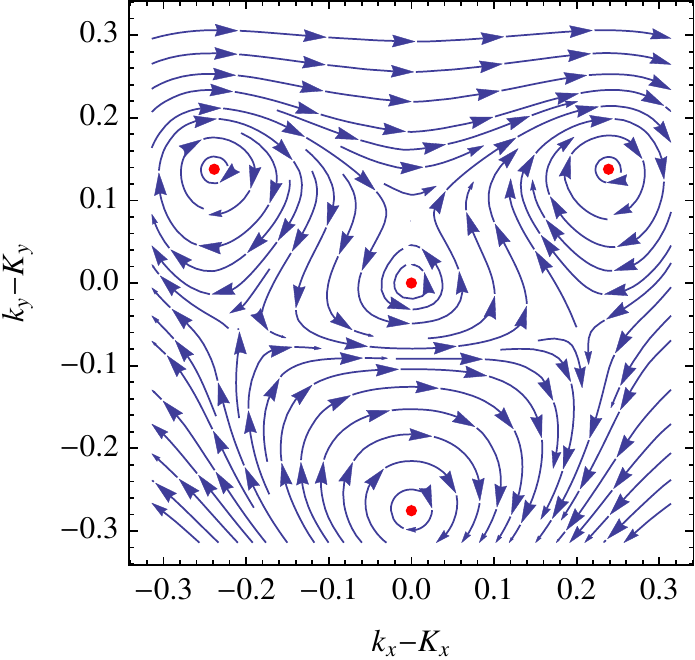}
\caption{Circulation of $\nabla\chi(\kv)$ around the Dirac point $\Kv$ and the new, satellite Dirac points, at $t'=0.4$.}      
\label{fig:chi}
\end{center}
\end{figure}

\begin{figure}[ht]
\begin{center}
\includegraphics[width = \hsize]{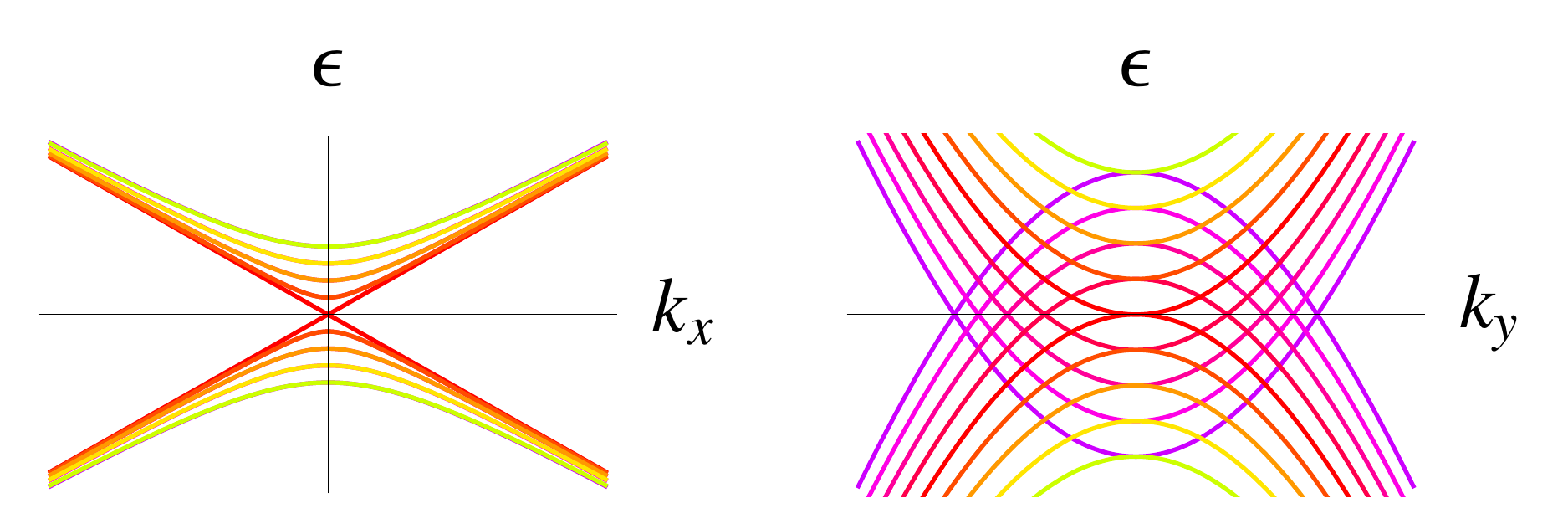}
\caption{(color online). 
Profile of the dispersion around $\frac12\Gv=(\Kv+\Kv')/2$, one of the merging locations of the new Dirac points at $t'=t'_c$.
The dispersion is shown along the $x$ (left) and $y$ (right) directions.
The various curves correspond to an array of values of $t'$ close to, and around $t'_c$.
Precisely at $t'_c$, the dispersion is linear in $k_x$ and quadratic in $k_y$.
For $t'\ne t'_c$' the two Dirac points stand away from $\frac12\Gv$.
}
\label{fig:merging}
\end{center}
\end{figure}

The integral of the PB curvature over occupied wave vectors is the Chern number. 
At half-filling, the two lowest-energy bands contribute 1 and $-1$, respectively, to the Chern number, coming from the first term of \eqref{eq:PB}.
Each Dirac point will in addition contribute $\pm\frac12$ to the Chern number, coming from the
second term of \eqref{eq:PB}, i.e., from the circulation of $\nabla\chi/4\pi$ around it.  
Figure~\ref{fig:chi} shows this circulation around the old and new Dirac points in the vicinity of $\Kv$.
From that figure, it appears clearly that the graphene Dirac point $\Kv$ will contribute $\frac12$, whereas the new Dirac points have the opposite contribution; but this picture is reversed when looking at the Dirac points surrounding $\Kv'$.
Thus the Dirac point contributions sum up to zero over the whole Brillouin zone, since they occur in pairs with opposite chiralities.

At half-filling, there is particle-hole symmetry and the explicit time-reversal breaking in the Hamiltonian does not induce any topology, because of zero total Chern number. We can think of this as an accidental restoration of time-reversal symmetry.

Note that the $t'=0$ line (the graphene limit) is singular in this respect. 
At $t'=0$, the bottom two bands become degenerate, and likewise for the top two bands.
Thus the graphene Dirac points $\Kv$ and $\Kv'$ arise for each of the two spin bands, and the two spins make exactly opposite contributions to the PB curvature. The latter is thus identically zero everywhere, whereas for $t'>0$ the PB curvature is not zero, but its integral over the Brillouin zone (the Chern number) is.

\subsection{Lifshitz transition}

At the critical value $t'_c$, the new Dirac points with opposite chiralities annihilate pairwise, at wave vectors $\frac12\Gv$ lying midway between zone corners. These merging wave vectors, indicated by green crosses on Fig.~\ref{fig:dirac}, are time-reversal invariant, since $-\frac12\Gv$ is equivalent to $\frac12\Gv$ because $\Gv$ is a reciprocal lattice vector.
This topological phase transition cannot be described by the total Chern number, which does not change here. 
It is a Lifshitz transition, akin to what has been described in Refs~\onlinecite{Lim:2012fk, Gail:2012uq, Yamaji:2006vn}.
Precisely at this transition, the dispersion around the merged Dirac points is linear in one direction and quadratic in the other, a behavior qualified as  \textit{semi-Dirac} in Ref.~\onlinecite{Gail:2012uq} and illustrated on Fig.~\ref{fig:merging}.
Experimentally, this transition could be probed by changes in the tunneling probability between the valence and conduction bands during Bloch-Zener oscillations.\cite{Lim:2012fk} 

\section{Interactions}\label{sec:interactions}

\subsection{The cluster dynamical impurity approximation}

In the interacting case, the Hamiltonian~\eqref{eq:H} must be treated within some approximation method. In this work, we use the cluster dynamical impurity approximation (CDIA), \cite{Balzer:2009kl, Senechal2010} closely related to the variational cluster approximation (VCA) and to the cellular dynamical mean field theory (CDMFT),\cite{Lichtenstein:2000vn,*Kotliar:2001} but more accurate in its rendering of the Mott transition. These methods can be understood in the framework of Potthoff's self-energy functional approach (SFA).\cite{Potthoff:2003b, [{For a recent review, see}][]Potthoff:2012fk}
In this approach, the physical self-energy of the system is obtained via a dynamical variational principle, expressed by the Euler equation
\begin{equation}\label{eq:euler}
\frac{\delta\Omega[\Sigmav]}{\delta\Sigmav} = 0
\end{equation}
The self-energy functional $\Omega[\Sigmav]$ is defined as follows:
\begin{equation} \label{eq:Omega1}
\Omega[\Sigmav]= F[\Sigmav]+\Tr\ln[-(\Gv^{-1}_0 -\Sigmav)^{-1}]
\end{equation}
where $\Gv_0$ is the Green function of the non-interacting part of Hamiltonian (\ref{eq:H}), 
and $F[\Sigmav]= \Phi[G[\Sigmav]]-\Tr(\Sigmav G[\Sigmav])$ is the Legendre transform 
of the Luttinger-Ward functional $\Phi[\Gv]$,\cite{Luttinger:1960uq} defined as a functional of the Green function $\Gv$.
We use a matrix notation for the Green function and self-energy to emphasize that they are to be considered as matrices in the space of frequencies and degrees of freedom (e.g. sites and spin). The symbol $\Tr$ means a functional trace, i.e a sum over bands, wave vectors and frequency.
At the stationary point of $\Omega[\Sigmav]$, the value of the functional $\Omega$ coincides with the thermodynamic grand potential $E-\mu N$.

\begin{figure}[ht]
\begin{center}
\includegraphics[scale=0.9]{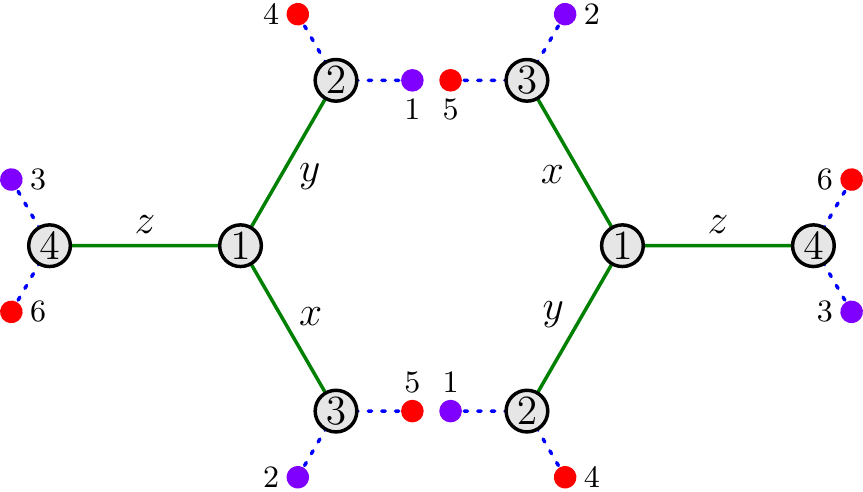}
\caption{(color online). The two-cluster system used in this work as a unit cell. Shaded numbered circles are lattice sites and small numbered circles represent the bath orbitals. The second cluster is a spatial inversion of the first.}
\label{fig:h4-6b}
\end{center}
\end{figure}

Unfortunately, the Luttinger-Ward functional $\Phi[\Gv]$, and consequently its Legendre transform $F[\Sigmav]$, is not known explicitly.
This leads to some approximations, in the weak-coupling regime, where $\Phi[\Gv]$ is represented by a truncated sum of diagrams.
The Hartree-Fock approximation is an example of such a truncation.
The basic idea behind the SFA is that the 
functional $F[\Sigmav]$ is an universal functional of the self-energy.\cite{Potthoff:2003b, Potthoff:2003} 
This means that the functional form of $F[\Sigmav]$ is the same for a 
reference Hamiltonian $H'$ with the same interaction as $H$, but a different
non interacting part. 
Typically, $H'$ will be a small system, e.g., a cluster, whose solution is known numerically.
Given the physical self-energy $\Sigmav(h)$ for a family of reference Hamiltonians $H'$ parametrized by $h$,
 the value of $F[\Sigmav(h)]$ can be extracted from the known grand potential $\Omega'[\Sigmav(h)]$ 
of these solutions, and thus the full Potthoff functional can be expressed as: 
\begin{multline}\label{eq:omega3}
 \Omega[\Sigmav(h)]=\Omega'[\Sigmav(h)]\\ +\Tr\ln[-(\Gv^{-1}_0 -\Sigmav(h))^{-1}]
 -\Tr\ln(-\Gv'(h))
\end{multline}
where $\Gv'$ is the known physical Green function of the reference system.
This relation provides us with an exact value of the functional $\Omega[\Sigmav(h)]$, albeit 
on a restricted space of self-energies $\Sigmav(h')$ which are the physical self-energies of 
the reference Hamiltonian $H'$.
If we introduce the notation $\Vv(\omega) = \Gv^{-1}_0-\Gv^{\prime-1}_0$, we can rewrite the above as
\begin{equation}\label{eq:omega4}
\Omega[\Sigmav(h)]=\Omega'[\Sigmav(h)] +\Tr\ln[1-\Vv(\omega)\Gv'(\omega)]
\end{equation}

Generally, the reference Hamiltonian $H'$ is based on a finite, periodically repeated cluster or set of clusters. This periodicity defines a superlattice, and a corresponding reduced Brillouin zone, smaller than the original lattice's Brillouin zone. The reference Green function $\Gv'$ is then momentum independent, and the matrix $\Vv$ depends on a single momentum (i.e., is diagonal in momentum indices).
Eq.~\eqref{eq:omega4} then reduces to the more explicit form
\begin{equation} \label{eq:omega5}
\Omega(h)=\Omega'(h) +\frac1N\sum_\kvt\ln\det[\mathbf{1}-\Vv(\kvt,\omega)\Gv'(\omega)]
\end{equation}
where now the matrices are `small', i.e., their order is the number of degrees of freedom in the repeated unit and $N$ is the (potentially large) number of lattice sites.

In the VCA, variational fields are added within the clusters, in order to give room to possible broken symmetries.
By contrast, CDMFT does not add extra terms to the cluster, but instead uses a set 
of non-interacting, fictitious orbitals (the bath) that are hybridized to the cluster and represent 
its immediate physical environment.
In CDMFT, the Potthoff functional \eqref{eq:omega3} is not calculated, and the solution is found 
instead by imposing a self-consistency relation between the cluster Green function $\Gv'$ and the 
projection of the lattice Green function $\Gv$ onto the cluster. In CDIA, a bath system is introduced, 
just like in CDMFT, but the solution is found by solving the Euler equation \eqref{eq:euler}, 
like in VCA. This allows us to also introduce variational fields on the cluster if needed,
 and at the same time gives a better description of temporal fluctuations (because of the 
presence of a bath), which is important to correctly capture the Mott transition.

\begin{figure}[ht]
\begin{center} 
\includegraphics[width=\hsize]{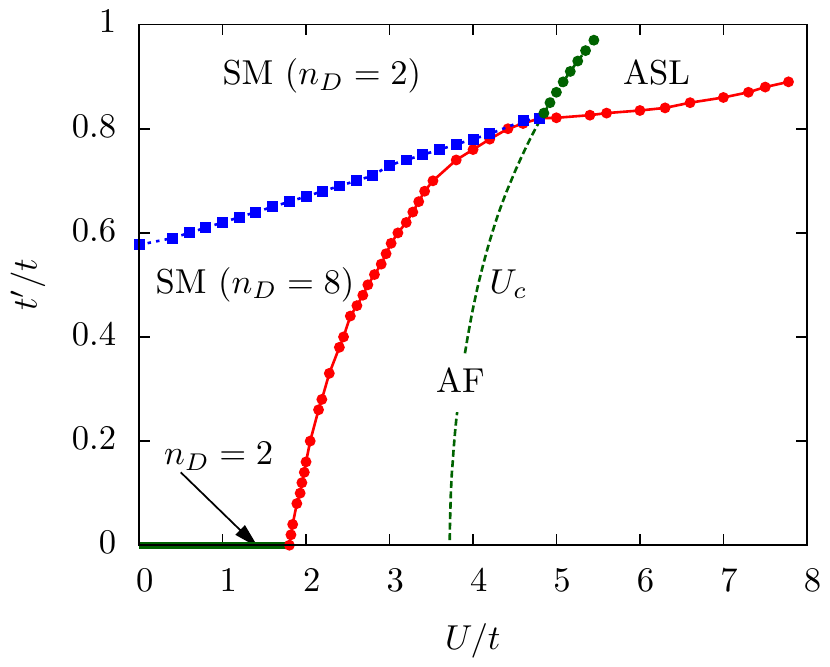}
\caption{(color online). Phase diagram of the half-filled Kitaev-Hubbard model (Eq.~(\ref{eq:H})) on the $U-t'$ plane ($t=1$). $n_D$ is to the  number of Dirac points in the semi-metallic (SM) phase; the $t'=0$ limit corresponds to graphene ($n_D=2$), represented by a green line. 
The dashed line indicates the first-order Mott transition in the normal (non magnetic) state.
The antiferromagnetic (AF) insulator phase is bounded by the red curve. The algebraic spin liquid (ALS) phase is the region of gapped spectrum and zero staggered magnetization.}
\label{fig:phasediagram}
\end{center}
\end{figure}
\begin{figure}[ht]
\begin{center} 
\includegraphics[width=0.9\hsize]{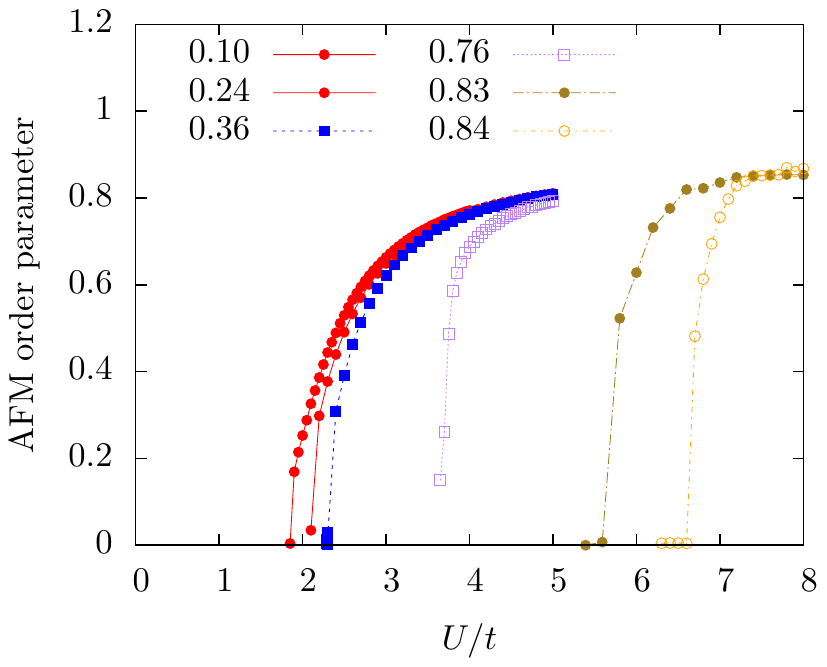}
\caption{(color online) The antiferromagnetic order parameter as a function of interaction $U$ 
for different spin-dependent hopping $t'$. We observe a continuous transition for $0<t'\leq0.4$, a weakly discontinuous transition when $0.4<t'<0.83$ and a discontinuous transition for $t'\geq0.83$.}
\label{fig:AF}
\end{center}
\end{figure}

\subsection{Reference system}

The reference system used in this work is based on a unit cell made of two four-site clusters, the second obtained from the first by a spatial inversion and a shift, as illustrated on Fig.~\ref{fig:h4-6b}. 
Together, these two clusters form an 8-site supercluster that tiles the honeycomb lattice.
Each of these two clusters contains four spatial sites and 6 bath sites.
The bath sites have no position per se, but can be thought of as representing the nearest sites of each cluster's environment, and are illustrated with this in mind on Fig.~\ref{fig:h4-6b} (small colored circles). This is why they are hybridized with the cluster boundary sites only, not the central site. 

The reference Hamiltonian has the following expression:
\begin{align}\label{eq:H'}
 H' &= \sum_{\langle i,j\rangle_\alpha} \left\{c^{\dag}_{i}\left(\frac{t+t'
\sigma^{\alpha}}{2}\right)c_{j}+{\rm H.c.}\right\} 
 +U \sum_in_{i\uparrow}n_{i\downarrow} \notag \\
&~~+\sum_{\mu,\sigma}\epsilon_\mu a^{\dag}_{\mu\sigma}a_{\mu\sigma}
+\sum_{i,\mu} c^{\dag}_{i}\left(\theta_{i\mu}+\vartheta_{i\mu}
\sigma^{\alpha(i,\mu)}\right)a_{\mu}+{\rm H.c.}
\end{align}
where $a_{\mu\sigma}$ annihilates an electron of spin $\sigma$ at the bath orbital $\mu$, $\epsilon_{\mu}$ is the energy of the bath orbital $\mu$, $\theta_{i\mu}$ 
is the hybridization between the bath orbital $\mu$ and site $i$, and $\vartheta_{i\mu}$ a corresponding spin-dependent hybridization.
The Pauli matrix $\sigma^{\alpha(i,\mu)}$ appearing in the bath hybridization is determined by the corresponding orientation ($x$, $y$ or $z$) of the hybridization link on Fig.~\ref{fig:h4-6b}.
$\epsilon_{\mu}$, $\theta_{i\mu}$ and $\vartheta_{i\mu}$ will be treated like variational parameters, and the solution adopted will be such that $\Omega(\epsilon,\theta)$ is stationary. Here the sum over $i,j$ is restricted to the cluster.
At a particle-hole symmetric point, such as the normal phase at $\mu=U/2$, only two of these bath parameters are independent, because of the symmetry of the cluster: $\theta_{i\mu} = \theta$, for all $(i,\mu)$ and $\epsilon_\mu = \pm\epsilon$, where the $+$ sign applies to $\mu=1,2,3$ and the $-$ sign to $\mu=4,5,6$. Particle-hole symmetry forces $\vartheta_{i\mu}$ to vanish; however, this will no longer be the case in the antiferromagnetic phase.

We probe the antiferromagnetic phase by adding to the cluster Hamiltonian $H'$ the term:
\begin{equation} \label{eq:M}
 H'_M = M\left[\sum_{i\in A}(n_{i,\uparrow}-n_{i,\downarrow})-
\sum_{i\in B}(n_{i,\uparrow}-n_{i,\downarrow})\right]
\end{equation}
where $n_{i,\sigma}=c^{\dag}_{i,\sigma}c_{i,\sigma}$,  
$A$ and $B$ stand for the two sublattices of the honeycomb lattice, and $M$, the antiferromagnetic Weiss field, is an  additional variational parameter.
The values of this Weiss field on the two clusters will be opposite.
In addition, as mentioned above, a spin-dependent hybridization $\vartheta_{i\mu} = \vartheta$, for all $(i,\mu)$, will be allowed, which makes a total of 4 independent variational parameters in that phase.

The matrix $\Vv(\kvt,\omega)$ of Eq.~\eqref{eq:omega5} contains all the information about the dispersion relation on the honeycomb lattice, including the hopping terms between the two clusters forming the repeated unit, as well as the hybridization functions associated with the baths connected to the two clusters. Because of the two clusters in the unit cell, all matrices ($\Gv'$, $\Vv$, etc.) have a block structure. The cluster Green function $\Gv'$ is block diagonal, but $\Vv$ isn't. However, all the frequency dependence of $\Vv$ lies in the block-diagonal components, and all the momentum dependence lies in the block off-diagonal components (that is not a general statement, but true for the system under study).

\subsection{Limitations of CDIA and cluster methods}

The strength of CDIA, and of other cluster methods like VCA, CDMFT and DCA, resides in their inclusion of short-range spatial correlations and of dynamical correlations. However, they have the following limitations:
(1) They do not take into account long range, two-particle fluctuations. Therefore they are insensitive to a possible destabilization of order by collective excitations, and in particular do not contain the physics behind the Mermin-Wagner theorem.
(2) Like mean-field theory, they cannot find orderings that are not programmed into them. Specifically, the bath parameters or Weiss fields must allow for a given broken symmetry to occur in order for the corresponding order to possibly emerge.
(3) The order probed must be commensurate with the repeated unit (unit cell) of the system; incommensurate order and order with large periods cannot be decribed in this framework.
(4) Anything about two-particle excitations is confined to the cluster itself, and suffers from strong finite-size effects.

Thus CDIA will make statements about the Mott transition or static (e.g. magnetic) order, but not about the type of spin liquid associated with the Mott phase. For that, other techniques must be used, as done in Ref.~\onlinecite{Hassan:2013fk}.
Even if we were to compute the dynamical spin susceptibility $\chi_{ij}(\omega)$, it would be confined to each cluster, and would show a sizeable spin gap due to finite-size effects alone, which would lead us to the (wrong) conclusion that the spin liquid associated with the Mott insulator is short-ranged instead of algebraic.

Both CDMFT and CDIA introduce bath orbitals to better capture quantum fluctuations in the time domain.
An advantage of CDIA over CDMFT lies in the possibility of adding Weiss fields to the cluster in order to probe broken symmetry phases more easily. Other advantages, described for instance in Refs~\onlinecite{Senechal2010,Potthoff:2012fk,Hassan:2013uq}, include a better description of the Mott transition (with a clear hysteresis between the metallic and insulating solutions). In addition, CDMFT with a finite bath has an ambiguity in its self-consistency procedure, which does not exist in CDIA, since the latter is based on an exact variational principle. However, for a given cluster and bath, CDIA is more demanding numerically, essentially because applying the variational principle requires a longer sequence of exact diagonalizations than the CDMFT self-consistent procedure, and convergence is more delicate.

\begin{figure}[ht]
\begin{center} 
\includegraphics[width=0.4\textwidth]{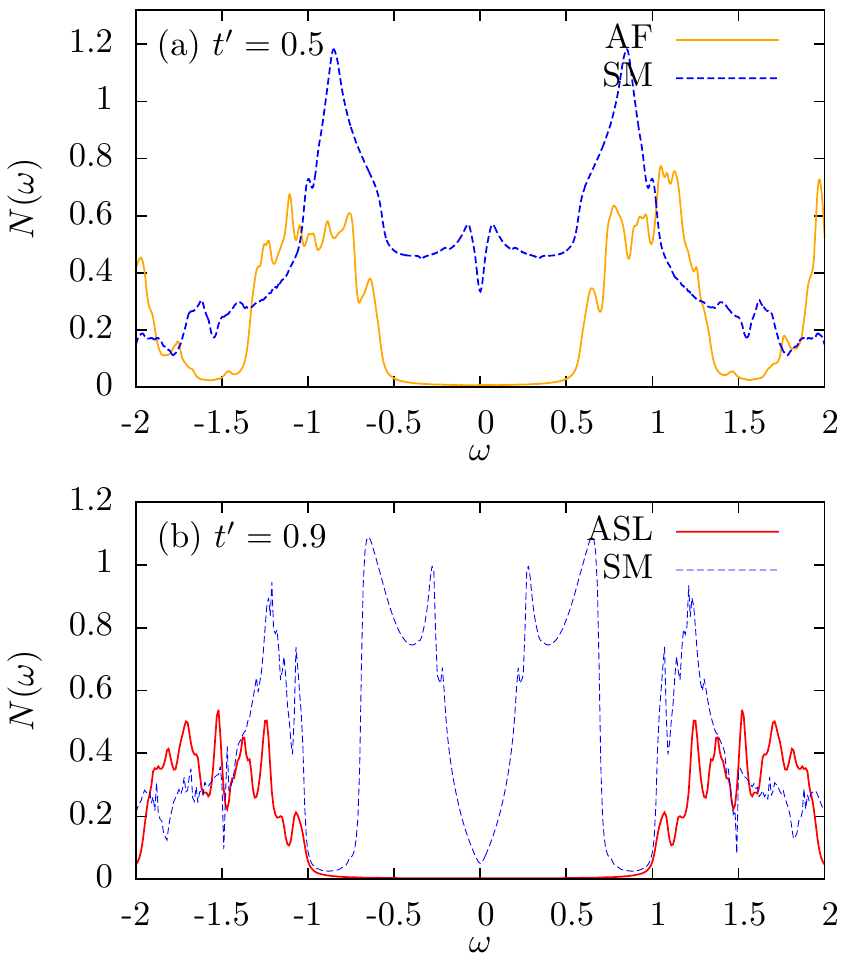}
\caption{(color online) Density of states $N(\omega)$ for different solutions on the phase diagram of Fig.~\ref{fig:phasediagram}. 
Top panel: $t'=0.5$; bottom panel: $t'=0.9$.
Note that the two semi-metallic solutions, ($t'=0.5t$, $U=t$) and ($t'=0.9t$, $U=t$), should have 
a vanishing density of states at the Fermi level, but  this is hidden here by the use of a Lorenzian broadening $\eta=0.01t$.
The semi-metal with eight Dirac points (top panel) has a narrower gap-like feature near the Fermi 
level, compared to the semi-metal with two Dirac points only (bottom panel).
}     
\label{fig:dos}
\end{center}
\end{figure}

\section{Results and discussion}\label{sec:results}

We first used the CDIA to locate the Mott transition as $U$ is increased from zero, for values of $t'$ in the interval $[0,1]$ (negative values of $t'$ are equivalent to positive values, except for the chirality of Dirac points, which is reversed).
We forbid the antiferromagnetic solution by setting the corresponding Weiss field to zero.
The Mott transition is discontinuous, displays hysteresis, and occurs at a critical value $U_c$ at which the grand potential $\Omega$ of the semi-metallic is the same as that of the insulating solutions. 
This transition line is shown as a green (dashed) line on Fig.~\ref{fig:phasediagram}.
When comparing with previous results on the Mott transition in graphene,\cite{Hassan:2013uq} one must recall the factor of $\frac12$ appearing in front of $t$ in the Hamiltonian \eqref{eq:H}, which means that the scale of the $U$ axis on Fig.~\ref{fig:phasediagram} must be multiplied by 2.

The semi-metallic side of the Mott transition is made of two different phases, depending on the number of distinct Dirac points (2 or 8).
This point was overlooked in Ref.~\onlinecite{Hassan:2013fk}.
At $U=0$, the transition between the two SM occurs at $t'= 1/\sqrt{3}$, as explained above.
For $U>0$, the transition is still visible by carefully looking at the spectral function computed from the CDIA solution and is indicated by the blue squares on Fig.~\ref{fig:phasediagram}.
Along the Mott line, this transition occurs towards $t'\approx 0.82$.
Note that the graphene limit ($t'=0$, indicated by a green full line on the figure) is singular, since the additional Dirac points, as well as the Berry curvature, appear as soon as $t'\ne 0$.

We then relax the constraint on the AF Weiss field $M$ to allow for long-range AF order and find the AF transition line shown as red squares on Fig.~\ref{fig:phasediagram}.
For $t'<0.82t$, the AF transition preempts the Mott transition and the spin liquid state does not
exist.
But since the spin-dependent hopping frustrates N\'eel order, the critical value $U_{\rm AF}$ recedes towards higher values as $t'$ increases, revealing the underlying spin liquid phase when $t'\geq 0.82t$.

\begin{figure}[ht]
\begin{center} 
\includegraphics[width=\hsize]{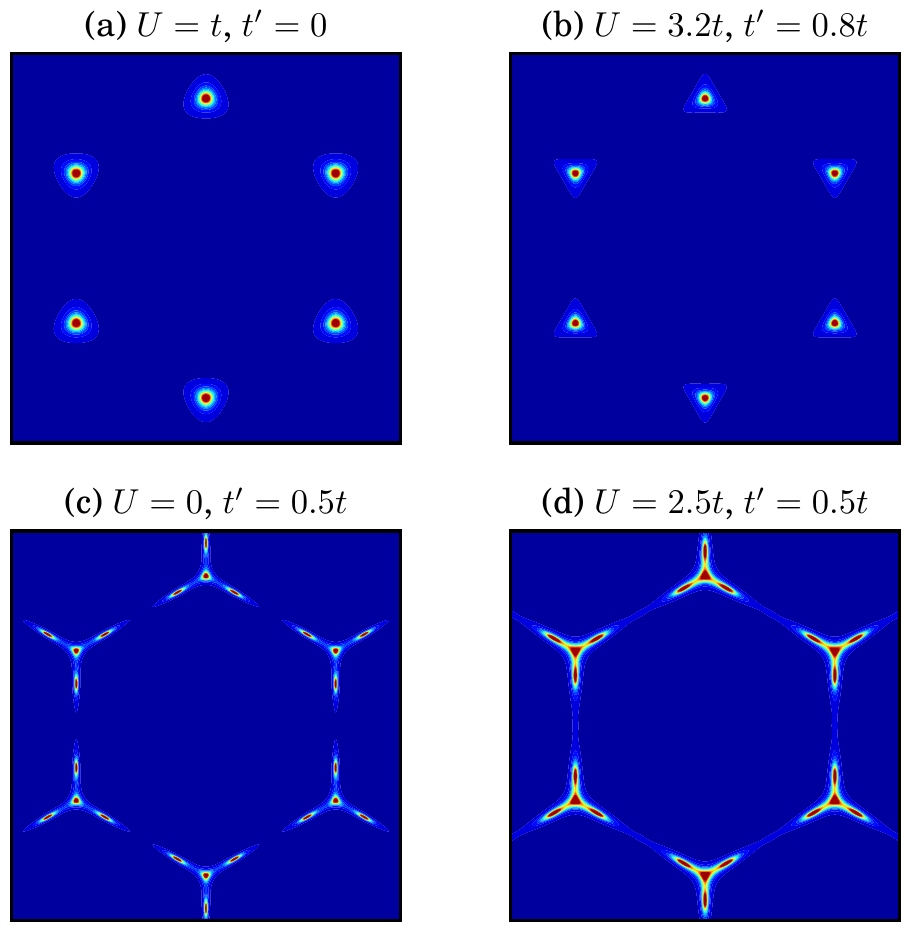}
\caption{(color online) Spectral function $A(\mathbf{k},0)$ at the Fermi level for different
semi-metallic solutions in the phase diagram (Fig.~\ref{fig:phasediagram}). Top left: the graphene limit ($t'=0$) at $U=t$, with two distinct Dirac points at $\mathbf{K}$ and $\mathbf{K}'$. Top right: the same, in the semi-metallic solution bordering on the ASL. On the bottom panel, the presence of 6 additional distinct Dirac points is clearly visible at $t'=0.5$, for both  non-interacting ($U=0$) and interacting ($U=2.5t$) solutions.
The Lorenzian broadening is $\eta=0.04t$ for panel (a) and $\eta = 0.01t$ for the others.}
\label{fig:mdc}
\end{center}
\end{figure}
Fig.~\ref{fig:AF} shows the behavior of the N\'eel order parameter as a function of $U$ for different values of the spin-dependent hopping $t'$.
For $t^{\prime}\leq 0.4t$ the order parameter
behaves as a square root (critical exponent of $\beta=1/2$) around the critical Coulomb repulsion $U_c$. 
This mean-field behavior occurs because cluster methods do not capture long wavelength fluctuations needed to correctly predict critical exponents.
When $t'$ increases ($0.4t<t^{\prime}<0.82t$), the transition becomes more abrupt and the square root behavior disappears; we call this a weakly discontinuous transition.
If $t'\geq 0.82t$, we observe clearly a jump of the order parameter which is a signature of a discontinuous phase transition between the AF and Mott (ASL) phases.
The discontinuous character of the transition is also seen when looking at the bath parameters, which show a clear jump at the transition.

Along the boundary between the semi-metal and the AF phase, we observe that both the graphene Dirac points and the new Dirac points disappear at once.
As soon as one enters the antiferromagnetic phase, the Weiss field $M$ is nonzero. This parameter breaks parity, and it is precisely that symmetry that protects the graphene Dirac points. 
If the Weiss field $M$ were part of a noninteracting Hamiltonian, then all Dirac point would disappear simultaneously, as shown around Eq.~\ref{eq:M0}. Thus it is not unnatural for the two types of Dirac points to disappear together.
Here the AF gap created at the Dirac points is a self-energy effect, but we should not be surprised that it affects all Dirac points simultaneously, in view of the $U=0$ behavior when parity is broken. In addition, particle-hole symmetry is broken as soon as we enter the AFM phase. Although we are lucky enough to be positioned within the AFM gap by chosing $\mu=U/2$, the spectral function is no longer symmetric around the Fermi level.

It is remarkable that the intersection of the N\'eel curve with the Mott curve coincides with the topological transition between the two types of semi-metals, even though the numerical procedures to determine that point are different. 
Thus the four phases identified in this work meet at $(U,t') \approx (4.8,0.82)$.
 
We will not argue here why the Mott phase found at $t'>0.82$ is an algebraic spin liquid.
The argument cannot be made using CDIA results, and can be found in Ref.~\onlinecite{Hassan:2013fk} and the associated supplementary material.
Note however that this phase appears at stronger Coulomb repulsion ($U/t>4.8$) than in Ref.~\onlinecite{Hassan:2013fk}, because of our use of CDIA instead of the simpler Cluster Perturbation Theory. 

Fig.~\ref{fig:dos} illustrates the density of states for AF insulator, ASL and SM regions of the phase diagram (see Fig.~\ref{fig:phasediagram}). The gap in the ASL and AF Mott insulator phase can be seen clearly at the Fermi level. For the same value $U=t$, the two semi-metallic phases have different low-energy structures, and each transits to a different phase upon increasing $U$.

In Fig.~\ref{fig:mdc}, we  represent the spectral function $A(\mathbf{k},\omega=0)$ at the Fermi level for different semi-metallic solutions of the phase diagram (Fig.~\ref{fig:phasediagram}).
The non-interacting case solved analytically at the beginning  is correctly represented with the six additional Dirac points when  $t^{\prime}\leq t'_c$ as shown in  Fig.~\ref{fig:mdc} (c). Above this critical value  and at $t^{\prime}=0$, the system is graphene-like with two Dirac points (Fig.~\ref{fig:mdc} (a) and (b)).
The features observed in (d) are similar to those of the noninteracting case (c), but broader. 

\begin{figure}[ht]
\begin{center} 
\includegraphics[scale=1.1]{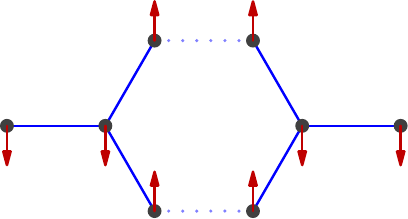}
\caption{(color online) Another possible magnetic order, with ordering wavevector $\Qv=\Mv$, probed in this work but for which no solution was found in the range of $U$ and $t'$ covered.\label{fig:stripy}}
\end{center}
\end{figure}

In principle, other magnetic orders could exist, and compete with both the simple AF order studied here and the spin liquid. We have probed a stripe-like collinear magnetic order with ordering wavevector $\Qv=\Mv$ (see Fig.~\ref{fig:stripy}), without finding any solution. Although this does not eliminate completely the possibility of competing order being present, it strengthens our point that the Mott phase exists when $t'$ is close to 1 and $U$ large enough.
 
\section{Conclusion}

We have investigated the phase diagram of the half-filled Kitaev-Hubbard model. 
The analytic solution in the non-interacting limit reveals a Lifshitz transition between two semi-metallic states, with two and eight Dirac points, respectively.
The Chern number is the same for these two phases, and the transition between the two occurs as the new Dirac points annihilate pairwise, forming a semi-Dirac point precisely at the transition.\cite{Gail:2012uq}
These two phases survive in the presence of interaction, as shown by an approximate solution of the interacting model using the Cluster Dynamical Impurity Approximation (CDIA).
In principle, the transition between the two semi-metals could be observed in Bloch-Zener tunneling\cite{Lim:2012fk} in a cold-atoms realization of the model.
Overall, the phase diagram contains four phases that meet at a single point.
On the strong coupling side, these are an antiferromagnetic phase at low $t'$, and a spin liquid phase (shown to be an algebraic spin liquid in Ref.~\onlinecite{Hassan:2013fk}) at high $t'$.
The transition from the antiferromagnet to the spin liquid is discontinuous, whereas the transition from the semi-metal to the antiferromagnet, which pre-empts a Mott transition, is continuous.

We gratefully acknowledge discussions with R.~Shankar and M.S.~Laad.
Computational resources were provided by Compute Canada and Calcul Qu\'ebec.


\begin{thebibliography}{29}%
\makeatletter
\providecommand \@ifxundefined [1]{%
 \@ifx{#1\undefined}
}%
\providecommand \@ifnum [1]{%
 \ifnum #1\expandafter \@firstoftwo
 \else \expandafter \@secondoftwo
 \fi
}%
\providecommand \@ifx [1]{%
 \ifx #1\expandafter \@firstoftwo
 \else \expandafter \@secondoftwo
 \fi
}%
\providecommand \natexlab [1]{#1}%
\providecommand \enquote  [1]{``#1''}%
\providecommand \bibnamefont  [1]{#1}%
\providecommand \bibfnamefont [1]{#1}%
\providecommand \citenamefont [1]{#1}%
\providecommand \href@noop [0]{\@secondoftwo}%
\providecommand \href [0]{\begingroup \@sanitize@url \@href}%
\providecommand \@href[1]{\@@startlink{#1}\@@href}%
\providecommand \@@href[1]{\endgroup#1\@@endlink}%
\providecommand \@sanitize@url [0]{\catcode `\\12\catcode `\$12\catcode
  `\&12\catcode `\#12\catcode `\^12\catcode `\_12\catcode `\%12\relax}%
\providecommand \@@startlink[1]{}%
\providecommand \@@endlink[0]{}%
\providecommand \url  [0]{\begingroup\@sanitize@url \@url }%
\providecommand \@url [1]{\endgroup\@href {#1}{\urlprefix }}%
\providecommand \urlprefix  [0]{URL }%
\providecommand \Eprint [0]{\href }%
\@ifxundefined \urlstyle {%
  \providecommand \doi  [0]{\begingroup \@sanitize@url \@doi}%
  \providecommand \@doi [1]{\endgroup \@@startlink {\doibase
  #1}doi:\discretionary {}{}{}#1\@@endlink }%
}{%
  \providecommand \doi  [0]{doi:\discretionary{}{}{}\begingroup
  \urlstyle{rm}\Url }%
}%
\providecommand \doibase [0]{http://dx.doi.org/}%
\providecommand \Doi [0]{\begingroup \@sanitize@url \@Doi }%
\providecommand \@Doi  [1]{\endgroup\@@startlink{\doibase#1}\@@Doi}%
\providecommand \@@Doi [1]{#1\@@endlink}%
\providecommand \selectlanguage [0]{\@gobble}%
\providecommand \bibinfo  [0]{\@secondoftwo}%
\providecommand \bibfield  [0]{\@secondoftwo}%
\providecommand \translation [1]{[#1]}%
\providecommand \BibitemOpen [0]{}%
\providecommand \bibitemStop [0]{}%
\providecommand \bibitemNoStop [0]{.\EOS\space}%
\providecommand \EOS [0]{\spacefactor3000\relax}%
\providecommand \BibitemShut  [1]{\csname bibitem#1\endcsname}%
\bibitem [{\citenamefont {Mott}(1968)}]{Mott:1968fk}%
  \BibitemOpen
  \bibfield  {author} {\bibinfo {author} {\bibfnamefont {N.~F.}\ \bibnamefont
  {Mott}},\ }\href {http://link.aps.org/doi/10.1103/RevModPhys.40.677}
  {\bibfield  {journal} {\bibinfo  {journal} {Rev. Mod. Phys.},\ }\textbf
  {\bibinfo {volume} {40}},\ \bibinfo {pages} {677} (\bibinfo {year}
  {1968})}\BibitemShut {NoStop}%
\bibitem [{\citenamefont {Imada}\ \emph {et~al.}(1998)\citenamefont {Imada},
  \citenamefont {Fujimori},\ and\ \citenamefont {Tokura}}]{Imada:1998rp}%
  \BibitemOpen
  \bibfield  {author} {\bibinfo {author} {\bibfnamefont {M.}~\bibnamefont
  {Imada}}, \bibinfo {author} {\bibfnamefont {A.}~\bibnamefont {Fujimori}}, \
  and\ \bibinfo {author} {\bibfnamefont {Y.}~\bibnamefont {Tokura}},\ }\Doi
  {10.1103/RevModPhys.70.1039} {\bibfield  {journal} {\bibinfo  {journal} {Rev.
  Mod. Phys.},\ }\textbf {\bibinfo {volume} {70}},\ \bibinfo {pages} {1039}
  (\bibinfo {year} {1998})}\BibitemShut {NoStop}%
\bibitem [{\citenamefont {Hassan}\ and\ \citenamefont
  {S{\'e}n{\'e}chal}(2013)}]{Hassan:2013uq}%
  \BibitemOpen
  \bibfield  {author} {\bibinfo {author} {\bibfnamefont {S.~R.}\ \bibnamefont
  {Hassan}}\ and\ \bibinfo {author} {\bibfnamefont {D.}~\bibnamefont
  {S{\'e}n{\'e}chal}},\ }\Doi {10.1103/PhysRevLett.110.096402} {\bibfield
  {journal} {\bibinfo  {journal} {Phys. Rev. Lett.},\ }\textbf {\bibinfo
  {volume} {110}},\ \bibinfo {pages} {096402} (\bibinfo {year}
  {2013})}\BibitemShut {NoStop}%
\bibitem [{\citenamefont {Fazekas}\ and\ \citenamefont
  {Anderson}(1974)}]{fazekas1974}%
  \BibitemOpen
  \bibfield  {author} {\bibinfo {author} {\bibfnamefont {P.}~\bibnamefont
  {Fazekas}}\ and\ \bibinfo {author} {\bibfnamefont {P.}~\bibnamefont
  {Anderson}},\ }\href@noop {} {\bibfield  {journal} {\bibinfo  {journal}
  {Philosophical Magazine},\ }\textbf {\bibinfo {volume} {30}},\ \bibinfo
  {pages} {423} (\bibinfo {year} {1974})}\BibitemShut {NoStop}%
\bibitem [{\citenamefont {Rantner}\ and\ \citenamefont
  {Wen}(2001)}]{Rantner2001}%
  \BibitemOpen
  \bibfield  {author} {\bibinfo {author} {\bibfnamefont {W.}~\bibnamefont
  {Rantner}}\ and\ \bibinfo {author} {\bibfnamefont {X.-G.}\ \bibnamefont
  {Wen}},\ }\href {http://link.aps.org/doi/10.1103/PhysRevLett.86.3871}
  {\bibfield  {journal} {\bibinfo  {journal} {Phys. Rev. Lett.},\ }\textbf
  {\bibinfo {volume} {86}},\ \bibinfo {pages} {3871} (\bibinfo {year}
  {2001})}\BibitemShut {NoStop}%
\bibitem [{\citenamefont {Affleck}\ and\ \citenamefont
  {Marston}(1988)}]{Affleck1988}%
  \BibitemOpen
  \bibfield  {author} {\bibinfo {author} {\bibfnamefont {I.}~\bibnamefont
  {Affleck}}\ and\ \bibinfo {author} {\bibfnamefont {J.~B.}\ \bibnamefont
  {Marston}},\ }\href {http://link.aps.org/doi/10.1103/PhysRevB.37.3774}
  {\bibfield  {journal} {\bibinfo  {journal} {Phys. Rev. B},\ }\textbf
  {\bibinfo {volume} {37}},\ \bibinfo {pages} {3774} (\bibinfo {year}
  {1988})}\BibitemShut {NoStop}%
\bibitem [{\citenamefont {Shimizu}\ \emph {et~al.}(2003)\citenamefont
  {Shimizu}, \citenamefont {Miyagawa}, \citenamefont {Kanoda}, \citenamefont
  {Maesato},\ and\ \citenamefont {Saito}}]{Shimizu:2003uq}%
  \BibitemOpen
  \bibfield  {author} {\bibinfo {author} {\bibfnamefont {Y.}~\bibnamefont
  {Shimizu}}, \bibinfo {author} {\bibfnamefont {K.}~\bibnamefont {Miyagawa}},
  \bibinfo {author} {\bibfnamefont {K.}~\bibnamefont {Kanoda}}, \bibinfo
  {author} {\bibfnamefont {M.}~\bibnamefont {Maesato}}, \ and\ \bibinfo
  {author} {\bibfnamefont {G.}~\bibnamefont {Saito}},\ }\href
  {http://link.aps.org/doi/10.1103/PhysRevLett.91.107001} {\bibfield  {journal}
  {\bibinfo  {journal} {Phys. Rev. Lett.},\ }\textbf {\bibinfo {volume} {91}},\
  \bibinfo {pages} {107001} (\bibinfo {year} {2003})}\BibitemShut {NoStop}%
\bibitem [{\citenamefont {Park}\ \emph {et~al.}(2003)\citenamefont {Park},
  \citenamefont {Park}, \citenamefont {Jeon}, \citenamefont {Choi},
  \citenamefont {Lee}, \citenamefont {Jo}, \citenamefont {Bewley},
  \citenamefont {McEwen},\ and\ \citenamefont {Perring}}]{Park:2003uq}%
  \BibitemOpen
  \bibfield  {author} {\bibinfo {author} {\bibfnamefont {J.}~\bibnamefont
  {Park}}, \bibinfo {author} {\bibfnamefont {J.-G.}\ \bibnamefont {Park}},
  \bibinfo {author} {\bibfnamefont {G.~S.}\ \bibnamefont {Jeon}}, \bibinfo
  {author} {\bibfnamefont {H.-Y.}\ \bibnamefont {Choi}}, \bibinfo {author}
  {\bibfnamefont {C.}~\bibnamefont {Lee}}, \bibinfo {author} {\bibfnamefont
  {W.}~\bibnamefont {Jo}}, \bibinfo {author} {\bibfnamefont {R.}~\bibnamefont
  {Bewley}}, \bibinfo {author} {\bibfnamefont {K.~A.}\ \bibnamefont {McEwen}},
  \ and\ \bibinfo {author} {\bibfnamefont {T.~G.}\ \bibnamefont {Perring}},\
  }\href {http://link.aps.org/doi/10.1103/PhysRevB.68.104426} {\bibfield
  {journal} {\bibinfo  {journal} {Phys. Rev. B},\ }\textbf {\bibinfo {volume}
  {68}},\ \bibinfo {pages} {104426} (\bibinfo {year} {2003})}\BibitemShut
  {NoStop}%
\bibitem [{\citenamefont {Helton}\ \emph {et~al.}(2007)\citenamefont {Helton},
  \citenamefont {Matan}, \citenamefont {Shores}, \citenamefont {Nytko},
  \citenamefont {Bartlett}, \citenamefont {Yoshida}, \citenamefont {Takano},
  \citenamefont {Suslov}, \citenamefont {Qiu}, \citenamefont {Chung},
  \citenamefont {Nocera},\ and\ \citenamefont {Lee}}]{Helton:2007ys}%
  \BibitemOpen
  \bibfield  {author} {\bibinfo {author} {\bibfnamefont {J.~S.}\ \bibnamefont
  {Helton}}, \bibinfo {author} {\bibfnamefont {K.}~\bibnamefont {Matan}},
  \bibinfo {author} {\bibfnamefont {M.~P.}\ \bibnamefont {Shores}}, \bibinfo
  {author} {\bibfnamefont {E.~A.}\ \bibnamefont {Nytko}}, \bibinfo {author}
  {\bibfnamefont {B.~M.}\ \bibnamefont {Bartlett}}, \bibinfo {author}
  {\bibfnamefont {Y.}~\bibnamefont {Yoshida}}, \bibinfo {author} {\bibfnamefont
  {Y.}~\bibnamefont {Takano}}, \bibinfo {author} {\bibfnamefont
  {A.}~\bibnamefont {Suslov}}, \bibinfo {author} {\bibfnamefont
  {Y.}~\bibnamefont {Qiu}}, \bibinfo {author} {\bibfnamefont {J.-H.}\
  \bibnamefont {Chung}}, \bibinfo {author} {\bibfnamefont {D.~G.}\ \bibnamefont
  {Nocera}}, \ and\ \bibinfo {author} {\bibfnamefont {Y.~S.}\ \bibnamefont
  {Lee}},\ }\href {http://link.aps.org/doi/10.1103/PhysRevLett.98.107204}
  {\bibfield  {journal} {\bibinfo  {journal} {Phys. Rev. Lett.},\ }\textbf
  {\bibinfo {volume} {98}},\ \bibinfo {pages} {107204} (\bibinfo {year}
  {2007})}\BibitemShut {NoStop}%
\bibitem [{\citenamefont {Lee}\ \emph {et~al.}(2007)\citenamefont {Lee},
  \citenamefont {Kikuchi}, \citenamefont {Qiu}, \citenamefont {Lake},
  \citenamefont {Huang}, \citenamefont {Habicht},\ and\ \citenamefont
  {Kiefer}}]{Lee:2007kx}%
  \BibitemOpen
  \bibfield  {author} {\bibinfo {author} {\bibfnamefont {S.-H.}\ \bibnamefont
  {Lee}}, \bibinfo {author} {\bibfnamefont {H.}~\bibnamefont {Kikuchi}},
  \bibinfo {author} {\bibfnamefont {Y.}~\bibnamefont {Qiu}}, \bibinfo {author}
  {\bibfnamefont {B.}~\bibnamefont {Lake}}, \bibinfo {author} {\bibfnamefont
  {Q.}~\bibnamefont {Huang}}, \bibinfo {author} {\bibfnamefont
  {K.}~\bibnamefont {Habicht}}, \ and\ \bibinfo {author} {\bibfnamefont
  {K.}~\bibnamefont {Kiefer}},\ }\href@noop {} {\bibfield  {journal} {\bibinfo
  {journal} {Nature materials},\ }\textbf {\bibinfo {volume} {6}},\ \bibinfo
  {pages} {853} (\bibinfo {year} {2007})}\BibitemShut {NoStop}%
\bibitem [{\citenamefont {Hermele}\ \emph {et~al.}(2008)\citenamefont
  {Hermele}, \citenamefont {Ran}, \citenamefont {Lee},\ and\ \citenamefont
  {Wen}}]{Hermele:2008vn}%
  \BibitemOpen
  \bibfield  {author} {\bibinfo {author} {\bibfnamefont {M.}~\bibnamefont
  {Hermele}}, \bibinfo {author} {\bibfnamefont {Y.}~\bibnamefont {Ran}},
  \bibinfo {author} {\bibfnamefont {P.~A.}\ \bibnamefont {Lee}}, \ and\
  \bibinfo {author} {\bibfnamefont {X.-G.}\ \bibnamefont {Wen}},\ }\href
  {http://link.aps.org/doi/10.1103/PhysRevB.77.224413} {\bibfield  {journal}
  {\bibinfo  {journal} {Phys. Rev. B},\ }\textbf {\bibinfo {volume} {77}},\
  \bibinfo {pages} {224413} (\bibinfo {year} {2008})}\BibitemShut {NoStop}%
\bibitem [{\citenamefont {Yan}\ \emph {et~al.}(2011)\citenamefont {Yan},
  \citenamefont {Huse},\ and\ \citenamefont {White}}]{Yan:2011fk}%
  \BibitemOpen
  \bibfield  {author} {\bibinfo {author} {\bibfnamefont {S.}~\bibnamefont
  {Yan}}, \bibinfo {author} {\bibfnamefont {D.~A.}\ \bibnamefont {Huse}}, \
  and\ \bibinfo {author} {\bibfnamefont {S.~R.}\ \bibnamefont {White}},\
  }\href@noop {} {\bibfield  {journal} {\bibinfo  {journal} {Science},\
  }\textbf {\bibinfo {volume} {332}},\ \bibinfo {pages} {1173} (\bibinfo {year}
  {2011})}\BibitemShut {NoStop}%
\bibitem [{\citenamefont {Iqbal}\ \emph {et~al.}(2013)\citenamefont {Iqbal},
  \citenamefont {Becca}, \citenamefont {Sorella},\ and\ \citenamefont
  {Poilblanc}}]{Iqbal:2013zr}%
  \BibitemOpen
  \bibfield  {author} {\bibinfo {author} {\bibfnamefont {Y.}~\bibnamefont
  {Iqbal}}, \bibinfo {author} {\bibfnamefont {F.}~\bibnamefont {Becca}},
  \bibinfo {author} {\bibfnamefont {S.}~\bibnamefont {Sorella}}, \ and\
  \bibinfo {author} {\bibfnamefont {D.}~\bibnamefont {Poilblanc}},\ }\href
  {http://link.aps.org/doi/10.1103/PhysRevB.87.060405} {\bibfield  {journal}
  {\bibinfo  {journal} {Phys. Rev. B},\ }\textbf {\bibinfo {volume} {87}},\
  \bibinfo {pages} {060405} (\bibinfo {year} {2013})}\BibitemShut {NoStop}%
\bibitem [{\citenamefont {Sahebsara}\ and\ \citenamefont
  {S{\'e}n{\'e}chal}(2008)}]{Sahebsara:2008fv}%
  \BibitemOpen
  \bibfield  {author} {\bibinfo {author} {\bibfnamefont {P.}~\bibnamefont
  {Sahebsara}}\ and\ \bibinfo {author} {\bibfnamefont {D.}~\bibnamefont
  {S{\'e}n{\'e}chal}},\ }\href@noop {} {\bibfield  {journal} {\bibinfo
  {journal} {Phys. Rev. Lett.},\ }\textbf {\bibinfo {volume} {100}},\ \bibinfo
  {pages} {136402} (\bibinfo {year} {2008})}\BibitemShut {NoStop}%
\bibitem [{\citenamefont {Tikhonov}\ \emph {et~al.}(2011)\citenamefont
  {Tikhonov}, \citenamefont {Feigel'man},\ and\ \citenamefont
  {Kitaev}}]{Tikhonov}%
  \BibitemOpen
  \bibfield  {author} {\bibinfo {author} {\bibfnamefont {K.~S.}\ \bibnamefont
  {Tikhonov}}, \bibinfo {author} {\bibfnamefont {M.~V.}\ \bibnamefont
  {Feigel'man}}, \ and\ \bibinfo {author} {\bibfnamefont {A.~Y.}\ \bibnamefont
  {Kitaev}},\ }\href {http://link.aps.org/doi/10.1103/PhysRevLett.106.067203}
  {\bibfield  {journal} {\bibinfo  {journal} {Phys. Rev. Lett.},\ }\textbf
  {\bibinfo {volume} {106}},\ \bibinfo {pages} {067203} (\bibinfo {year}
  {2011})}\BibitemShut {NoStop}%
\bibitem [{\citenamefont {Kitaev}(2006)}]{Kitaev:2006fk}%
  \BibitemOpen
  \bibfield  {author} {\bibinfo {author} {\bibfnamefont {A.}~\bibnamefont
  {Kitaev}},\ }\href
  {http://www.sciencedirect.com/science/article/pii/S0003491605002381}
  {\bibfield  {journal} {\bibinfo  {journal} {Annals of Physics},\ }\textbf
  {\bibinfo {volume} {321}},\ \bibinfo {pages} {2 } (\bibinfo {year} {2006})},\
  ISSN \bibinfo {issn} {0003-4916},\ \bibinfo {note} {january Special
  Issue}\BibitemShut {NoStop}%
\bibitem [{\citenamefont {Hassan}\ \emph
  {et~al.}(2013){\natexlab{a}}\citenamefont {Hassan}, \citenamefont
  {Sriluckshmy}, \citenamefont {Goyal}, \citenamefont {Shankar},\ and\
  \citenamefont {S{\'e}n{\'e}chal}}]{Hassan:2013fk}%
  \BibitemOpen
  \bibfield  {author} {\bibinfo {author} {\bibfnamefont {S.~R.}\ \bibnamefont
  {Hassan}}, \bibinfo {author} {\bibfnamefont {P.~V.}\ \bibnamefont
  {Sriluckshmy}}, \bibinfo {author} {\bibfnamefont {S.~K.}\ \bibnamefont
  {Goyal}}, \bibinfo {author} {\bibfnamefont {R.}~\bibnamefont {Shankar}}, \
  and\ \bibinfo {author} {\bibfnamefont {D.}~\bibnamefont {S{\'e}n{\'e}chal}},\
  }\href {http://link.aps.org/doi/10.1103/PhysRevLett.110.037201} {\bibfield
  {journal} {\bibinfo  {journal} {Phys. Rev. Lett.},\ }\textbf {\bibinfo
  {volume} {110}},\ \bibinfo {pages} {037201} (\bibinfo {year}
  {2013}{\natexlab{a}})}\BibitemShut {NoStop}%
\bibitem [{\citenamefont {Hassan}\ \emph
  {et~al.}(2013){\natexlab{b}}\citenamefont {Hassan}, \citenamefont {Goyal},
  \citenamefont {Shankar},\ and\ \citenamefont {S\'en\'echal}}]{Hassan:2013kx}%
  \BibitemOpen
  \bibfield  {author} {\bibinfo {author} {\bibfnamefont {S.~R.}\ \bibnamefont
  {Hassan}}, \bibinfo {author} {\bibfnamefont {S.}~\bibnamefont {Goyal}},
  \bibinfo {author} {\bibfnamefont {R.}~\bibnamefont {Shankar}}, \ and\
  \bibinfo {author} {\bibfnamefont {D.}~\bibnamefont {S\'en\'echal}},\ }\href
  {http://link.aps.org/doi/10.1103/PhysRevB.88.045301} {\bibfield  {journal}
  {\bibinfo  {journal} {Phys. Rev. B},\ }\textbf {\bibinfo {volume} {88}},\
  \bibinfo {pages} {045301} (\bibinfo {year} {2013}{\natexlab{b}})}\BibitemShut
  {NoStop}%
\bibitem [{\citenamefont {Lim}\ \emph {et~al.}(2012)\citenamefont {Lim},
  \citenamefont {Fuchs},\ and\ \citenamefont {Montambaux}}]{Lim:2012fk}%
  \BibitemOpen
  \bibfield  {author} {\bibinfo {author} {\bibfnamefont {L.-K.}\ \bibnamefont
  {Lim}}, \bibinfo {author} {\bibfnamefont {J.-N.}\ \bibnamefont {Fuchs}}, \
  and\ \bibinfo {author} {\bibfnamefont {G.}~\bibnamefont {Montambaux}},\
  }\href {http://link.aps.org/doi/10.1103/PhysRevLett.108.175303} {\bibfield
  {journal} {\bibinfo  {journal} {Phys. Rev. Lett.},\ }\textbf {\bibinfo
  {volume} {108}},\ \bibinfo {pages} {175303} (\bibinfo {year}
  {2012})}\BibitemShut {NoStop}%
\bibitem [{\citenamefont {de~Gail}\ \emph {et~al.}(2012)\citenamefont
  {de~Gail}, \citenamefont {Fuchs}, \citenamefont {Goerbig}, \citenamefont
  {Pi{\'e}chon},\ and\ \citenamefont {Montambaux}}]{Gail:2012uq}%
  \BibitemOpen
  \bibfield  {author} {\bibinfo {author} {\bibfnamefont {R.}~\bibnamefont
  {de~Gail}}, \bibinfo {author} {\bibfnamefont {J.-N.}\ \bibnamefont {Fuchs}},
  \bibinfo {author} {\bibfnamefont {M.}~\bibnamefont {Goerbig}}, \bibinfo
  {author} {\bibfnamefont {F.}~\bibnamefont {Pi{\'e}chon}}, \ and\ \bibinfo
  {author} {\bibfnamefont {G.}~\bibnamefont {Montambaux}},\ }\href@noop {}
  {\bibfield  {journal} {\bibinfo  {journal} {Physica B: Condensed Matter},\
  }\textbf {\bibinfo {volume} {407}},\ \bibinfo {pages} {1948} (\bibinfo {year}
  {2012})}\BibitemShut {NoStop}%
\bibitem [{\citenamefont {Yamaji}\ \emph {et~al.}(2006)\citenamefont {Yamaji},
  \citenamefont {Misawa},\ and\ \citenamefont {Imada}}]{Yamaji:2006vn}%
  \BibitemOpen
  \bibfield  {author} {\bibinfo {author} {\bibfnamefont {Y.}~\bibnamefont
  {Yamaji}}, \bibinfo {author} {\bibfnamefont {T.}~\bibnamefont {Misawa}}, \
  and\ \bibinfo {author} {\bibfnamefont {M.}~\bibnamefont {Imada}},\ }\href
  {http://jpsj.ipap.jp/link?JPSJ/75/094719/} {\bibfield  {journal} {\bibinfo
  {journal} {Journal of the Physical Society of Japan},\ }\textbf {\bibinfo
  {volume} {75}},\ \bibinfo {pages} {094719} (\bibinfo {year}
  {2006})}\BibitemShut {NoStop}%
\bibitem [{\citenamefont {Balzer}\ \emph {et~al.}(2009)\citenamefont {Balzer},
  \citenamefont {Kyung}, \citenamefont {S{\'e}n{\'e}chal}, \citenamefont
  {Tremblay},\ and\ \citenamefont {Potthoff}}]{Balzer:2009kl}%
  \BibitemOpen
  \bibfield  {author} {\bibinfo {author} {\bibfnamefont {M.}~\bibnamefont
  {Balzer}}, \bibinfo {author} {\bibfnamefont {B.}~\bibnamefont {Kyung}},
  \bibinfo {author} {\bibfnamefont {D.}~\bibnamefont {S{\'e}n{\'e}chal}},
  \bibinfo {author} {\bibfnamefont {A.~M.~S.}\ \bibnamefont {Tremblay}}, \ and\
  \bibinfo {author} {\bibfnamefont {M.}~\bibnamefont {Potthoff}},\ }\Doi
  {10.1209/0295-5075/85/17002} {\bibfield  {journal} {\bibinfo  {journal}
  {Europhys. Lett.},\ }\textbf {\bibinfo {volume} {85}},\ \bibinfo {pages}
  {17002} (\bibinfo {year} {2009})}\BibitemShut {NoStop}%
\bibitem [{\citenamefont {S\'en\'echal}(2010)}]{Senechal2010}%
  \BibitemOpen
  \bibfield  {author} {\bibinfo {author} {\bibfnamefont {D.}~\bibnamefont
  {S\'en\'echal}},\ }\href {http://link.aps.org/doi/10.1103/PhysRevB.81.235125}
  {\bibfield  {journal} {\bibinfo  {journal} {Phys. Rev. B},\ }\textbf
  {\bibinfo {volume} {81}},\ \bibinfo {pages} {235125} (\bibinfo {year}
  {2010})}\BibitemShut {NoStop}%
\bibitem [{\citenamefont {Lichtenstein}\ and\ \citenamefont
  {Katsnelson}(2000)}]{Lichtenstein:2000vn}%
  \BibitemOpen
  \bibfield  {author} {\bibinfo {author} {\bibfnamefont {A.~I.}\ \bibnamefont
  {Lichtenstein}}\ and\ \bibinfo {author} {\bibfnamefont {M.~I.}\ \bibnamefont
  {Katsnelson}},\ }\Doi {10.1103/PhysRevB.62.R9283} {\bibfield  {journal}
  {\bibinfo  {journal} {Phys. Rev. B},\ }\textbf {\bibinfo {volume} {62}},\
  \bibinfo {pages} {R9283} (\bibinfo {year} {2000})}\BibitemShut {NoStop}%
\bibitem [{\citenamefont {Kotliar}\ \emph {et~al.}(2001)\citenamefont
  {Kotliar}, \citenamefont {Savrasov}, \citenamefont {P{\'a}lsson},\ and\
  \citenamefont {Biroli}}]{Kotliar:2001}%
  \BibitemOpen
  \bibfield  {author} {\bibinfo {author} {\bibfnamefont {G.}~\bibnamefont
  {Kotliar}}, \bibinfo {author} {\bibfnamefont {S.Y.}~\bibnamefont {Savrasov}},
  \bibinfo {author} {\bibfnamefont {G.}~\bibnamefont {P{\'a}lsson}}, \ and\
  \bibinfo {author} {\bibfnamefont {G.}~\bibnamefont {Biroli}},\ }\href@noop {}
  {\bibfield  {journal} {\bibinfo  {journal} {Phys. Rev. Lett.},\ }\textbf
  {\bibinfo {volume} {87}},\ \bibinfo {pages} {186401} (\bibinfo {year}
  {2001})}\BibitemShut {NoStop}%
\bibitem [{\citenamefont {Potthoff}(2003)}]{Potthoff:2003b}%
  \BibitemOpen
  \bibfield  {author} {\bibinfo {author} {\bibfnamefont {M.}~\bibnamefont
  {Potthoff}},\ }\Doi {10.1140/epjb/e2003-00121-8} {\bibfield  {journal}
  {\bibinfo  {journal} {Eur. Phys. J. B},\ }\textbf {\bibinfo {volume} {32}},\
  \bibinfo {pages} {429 } (\bibinfo {year} {2003})}\BibitemShut {NoStop}%
\bibitem [{\citenamefont {Potthoff}(2012)}]{Potthoff:2012fk}%
  \BibitemOpen
  \bibfield  {author} {\bibinfo {author} {\bibfnamefont {M.}~\bibnamefont
  {Potthoff}},\ }in\ \href@noop {} {\emph {\bibinfo {booktitle} {Theoretical
  methods for Strongly Correlated Systems}}},\ \bibinfo {series} {Springer
  Series in Solid-State Sciences}, Vol.\ \bibinfo {volume} {171},\ \bibinfo
  {editor} {edited by\ \bibinfo {editor} {\bibfnamefont {A.}~\bibnamefont
  {Avella}}\ and\ \bibinfo {editor} {\bibfnamefont {F.}~\bibnamefont
  {Mancini}}}\ (\bibinfo  {publisher} {Springer},\ \bibinfo {year} {2012})\
  Chap.~\bibinfo {chapter} {9}\BibitemShut {NoStop}%
\bibitem [{\citenamefont {Luttinger}\ and\ \citenamefont
  {Ward}(1960)}]{Luttinger:1960uq}%
  \BibitemOpen
  \bibfield  {author} {\bibinfo {author} {\bibfnamefont {J.~M.}\ \bibnamefont
  {Luttinger}}\ and\ \bibinfo {author} {\bibfnamefont {J.~C.}\ \bibnamefont
  {Ward}},\ }\Doi {10.1103/PhysRev.118.1417} {\bibfield  {journal} {\bibinfo
  {journal} {Phys. Rev.},\ }\textbf {\bibinfo {volume} {118}},\ \bibinfo
  {pages} {1417} (\bibinfo {year} {1960})}\BibitemShut {NoStop}%
\bibitem [{\citenamefont {Potthoff}\ \emph {et~al.}(2003)\citenamefont
  {Potthoff}, \citenamefont {Aichhorn},\ and\ \citenamefont
  {Dahnken}}]{Potthoff:2003}%
  \BibitemOpen
  \bibfield  {author} {\bibinfo {author} {\bibfnamefont {M.}~\bibnamefont
  {Potthoff}}, \bibinfo {author} {\bibfnamefont {M.}~\bibnamefont {Aichhorn}},
  \ and\ \bibinfo {author} {\bibfnamefont {C.}~\bibnamefont {Dahnken}},\
  }\href@noop {} {\bibfield  {journal} {\bibinfo  {journal} {Phys. Rev.
  Lett.},\ }\textbf {\bibinfo {volume} {91}},\ \bibinfo {pages} {206402}
  (\bibinfo {year} {2003})}\BibitemShut {NoStop}%
\end{thebibliography}

%

\end{document}